\documentclass[acmsmall]{acmart}

\usepackage{multirow}
\usepackage{makecell}

\AtBeginDocument{%
  \providecommand\BibTeX{{%
    \normalfont B\kern-0.5em{\scshape i\kern-0.25em b}\kern-0.8em\TeX}}}

\setcopyright{acmcopyright}
\copyrightyear{0000}
\acmYear{0000}
\acmDOI{0/0}

\acmJournal{JACM}
\acmVolume{0}
\acmNumber{0}
\acmArticle{111}
\acmMonth{0}

\begin{document}

\title{Explaining Documents' Relevance to Search Queries}


\author{Razieh Rahimi}
\affiliation{%
  \institution{Center for Intelligent Information Retrieval, University of Massachusetts Amherst}
  \streetaddress{140 Governors Dr}
  \city{Amherst}
  \state{Massachusetts}
  \postcode{01002}
  \country{USA}
  }
\email{rahimi@cs.umass.edu}

\author{Youngwoo Kim}
\affiliation{%
  \institution{Center for Intelligent Information Retrieval, University of Massachusetts Amherst}
  \country{USA}
  }
\email{youngwookim@cs.umass.edu}

\author{Hamed Zamani}
\affiliation{%
  \institution{Center for Intelligent Information Retrieval, University of Massachusetts Amherst}
  \country{USA}
  }
\email{zamani@cs.umass.edu}

\author{James Allan}
\affiliation{%
  \institution{Center for Intelligent Information Retrieval, University of Massachusetts Amherst}
  \country{USA}
  }
\email{allan@cs.umass.edu}

\renewcommand{\shortauthors}{Rahimi, et al.}

\begin{abstract}

We present \textbf{GenEx}, a generative model to explain search results to users beyond just showing matches between query and document words. Adding GenEx explanations to search results greatly impacts user satisfaction and search performance. Search engines mostly provide document titles,  URLs,  and snippets for each result. Existing model-agnostic explanation methods similarly focus on word matching or content-based features. However, a recent user study shows that word matching features are quite obvious to users and thus of slight value. GenEx explains a search result by providing a terse description for the query aspect covered by that result. We cast the task as a sequence transduction problem and propose a novel model based on the Transformer architecture. To represent documents with respect to the given queries and yet not generate the queries themselves as explanations, two \emph{query-attention layers} and \emph{masked-query decoding} are added to the Transformer architecture. The model is trained without using any human-generated explanations. Training data are instead automatically constructed to ensure a tolerable noise level and a generalizable learned model. Experimental evaluation shows that our explanation models significantly outperform the baseline models. Evaluation through user studies also demonstrates that our explanation model generates short yet useful explanations.

\end{abstract}



\maketitle

\section{Introduction}

We focus on a new class of explanations for search results that aims to help users gain deeper understanding of search results and that is suitable for different search scenarios, from ad-hoc to conversational information seeking, and from desktop computers to voice-based systems. 

Search engine result pages currently provide document snippets in addition to document titles for each result, where the snippets are typically 2- or 3-line extracts highlighting the query words in the documents' contents. Although it is known that the quality of that summary can have a significant effect on user interactions~\cite{Mi:2019:UIS:3331184.3331306}, search snippets oftentimes are not coherent \cite{Tombros:1998:AQB:290941.290947} and fail to explain the documents' relevance to the submitted query~\cite{10.1145/3371390}. Thus the users may be confused as why a document is presented in  search results.

To address all these issues, we propose \textsc{GenEx}, an approach that generates terse explanations -- on the level of noun phrases -- for search results describing what aspect of the query is covered by a retrieved document.
For example, suppose that in response to the query ``\textsc{Obamacare}'' a document is listed that discusses how income is subject to an additional tax. A desired  explanation for the document  is ``\textsc{impacts  on medicare tax}'', which provides information beyond that of the snippet: ``\textsc{..tax to offset the costs of the OBAMACARE. this tax first took effect in 2013...}'' (as automatically generated by the Indri search engine).

We start this work by describing a set of studies to assess the usefulness of explanations like those produced by \textsc{GenEx} -- that is, to explore whether the proposed explanations can help users make more accurate and/or faster relevance decisions?  
In Section~\ref{sec:usability}, we describe and then compare two presentations of search results:~1)~showing documents' snippets only, and 2)~showing snippets and explanations.
We demonstrate that when participants have the explanations, they reach consensus on relevant document in 23\% more cases than when they have snippets alone. In addition, participants could detect the relevant document in 7 fewer seconds on average (22\% faster)  when explanations are provided.

The problem of constructing such explanations does not appear to have been studied previously in works on snippet generation or model-agnostic search result explanation. 
Prior work on model-agnostic explanation of information retrieval models  is  either \emph{local},  focusing on explaining individual rankings~\cite{Singh:2019:EES:3289600.3290620}, or \emph{global},  explaining the model behavior as a whole by training a simpler ``interpretable'' ranker, such as decision trees or a linear ranker~\cite{singh2018posthoc}. Both types of work  focus on explaining content-based features in ranking: which words or word-matching features  contributed more in the provided rankings. Although useful in some  cases, the recent study by Thomas et al.~\cite{10.1145/3371390} suggests that users of search systems benefit more from explanations describing documents' relevance \emph{beyond} word matching. 

Explaining documents' relevance  faces the major challenging step of extracting and conceptually representing the  query-related part(s) of the document content (possibly a small part of the document~\cite{TREC}) with respect to generally vague short-keyword queries. In addition, as with the simpler task of general document explanation (such as headline generation~\cite{rush-etal-2015-neural}), obtaining substantial amounts of manually-labeled training data is  often costly and time consuming. 

In Section~\ref{sec:model}, we cast the task of search result explanation as a sequence transduction problem, where an attention-based encoder-decoder architecture first provides a topic-focused contextual representation of a document and then generates desired explanations.
We specifically extend the Transformer architecture~\cite{Vaswani:2017:AYN:3295222.3295349} by introducing a query attention layer in the encoder to represent query-focused parts of documents. We then mask the query in the decoder to  generate coherent textual explanations of  information  in documents that  satisfy the user information need.

We propose solutions to automatically obtain training data from the Web to bypass the expensive human labeling process. Our model is thus trained with no manually labeled training data.
To build weakly-labeled training data with a noise level that a supervised encoder-decoder model can tolerate and that learns a model that generalizes to open-domain input texts, we combine samples from two sources: (1)~Wikipedia articles as more controlled edited content than the entire Web, and (2)~anchor texts in a collection of general web pages. 



In Section~\ref{sec:exp}, we describe extensive experiments on multiple datasets to evaluate \textsc{GenEx}.
Our results show that \textsc{GenEx} significantly outperforms the baselines: it improves BLEU-1 by 67\%-73\% as well as ROUGE-1 and ROUGE-L by 47\% - 51\%, all  over the original transformer architecture on general test samples from the Web.

Since BLEU and ROUGE do not always correlate with human judgments, we continue in Section~\ref{sec:humancheck} with a final study asking people to evaluate the quality of generated explanations, incorporating both relevance to the query and match to the documents' content.
The  results show that GenEx  explanations  are preferred over the strongest baseline by a majority of workers in 73\% of samples, and tied in another 7\%.


All datasets and user annotations collected in this study will be publicly available. 
\section{Related Work}
\label{sec:rel}
We review the related previous work on document summarization, snippet generation, and explainable search and recommendation.

 \subsection{Document Summarization} 
Document summarization is related to the defined task of search result explanation. As we aim to generate coherent and grammatically readable explanations, abstractive models for text summarization~\cite{rush-etal-2015-neural,liu-etal-2015-toward,nallapati-etal-2016-abstractive,chopra-etal-2016-abstractive} better suit as finishing components of explanation generation models, compared to extractive summarization models~\cite{rossiello-etal-2017-centroid}.
Neural abstractive summarization  started by generating headlines from the first sentence of  news articles~\cite{rush-etal-2015-neural} and was then applied to different settings, such as longer text inputs~\cite{liu-etal-2015-toward,nallapati-etal-2016-abstractive,chopra-etal-2016-abstractive}.
Although search result explanation is beyond summarizing document contents, even exploiting abstractive summarization techniques to different types of documents in the open-domain Web is not straightforward, as they need domain-specific fine-tuning  at the very least to produce reasonable outputs~\cite{pmlr-v97-chu19b}.



Dealing with no readily available labeled data, unsupervised training of sequence transduction tasks has been studied in some recent work.
Since language understanding is required to generate fluent sequences (texts), using pre-trained language representation models, such as Word2vec and GloVe embeddings~\cite{Mikolov:2013:DRW:2999792.2999959, pennington-etal-2014-glove}, ELMo~\cite{peters-etal-2018-deep},  BERT~\cite{devlin-etal-2019-bert}, and \textsc{UniLM}~\cite{dong2019unified} can reduce the amount of training data required. 
The pre-trained language representation model Bert is used for   ensuring fluency of generated text in the decoding step or representing 
 input   documents  in the encoding step  of text summarization models~\cite{zhang-etal-2019-pretraining,liu2019text}.
Beyond using pre-trained components in conjunction with or as part of a sequence-to-sequence model, there are some  
pre-trained models for the sequence generation task~\cite{pmlr-v97-song19d,lewis2019bart}. 
Unsupervised training of machine translation models as another example of sequence transduction has also been investigated in some studies~\cite{artetxe2018unsupervised,lample2019cross,ren-etal-2019-explicit,artetxe-etal-2019-effective}.




\subsection{Snippet Generation}
Snippet generation has been known a special type of  document summarization, in which sentences, or sentence fragments, are selected to be presented in a search engine result page (SERP)~\cite{Turpin:2007}.  It  was also called query-biased summarization
by \citet{Tombros:1998:AQB:290941.290947}. 
Snippet generation is an active area of research and has been studied in the context of Web search~\cite{verstak2012generation}, XML retrieval~\cite{Huang:2008}, semantic search~\cite{Wang:2019}, and more recently dataset search~\cite{10.1007/978-3-030-30793-6_39}.
Early Web search engines presented query-independent snippets consisting of the first tokens of the result document. 
Google was the first Web search engine to provide query-biased summaries~\cite{verstak2012generation,Turpin:2007}.
\citet{Bast:2014} proposed an efficient solution for extractive snippets by taking advantage of inverted index, a popular data structure used in most information retrieval systems.
Recently, \citet{Chen:2020} proposed abstractive snippet generation as a potential solution to circumvent copyright issues. The authors demonstrated that despite the popularity of extractive snippets in the current search engines, abstractive summarization is equally powerful in terms of user acceptance and expressiveness. 

Although snippet generation, and in general query-focused document summarization, is closely related to GenEx explanation, they are fundamentally different.
GenEx explanations are terse, consisting of a few words, while snippet generation models try to select or generate a few sentences or even a paragraph. In addition to the length, the goal of these  tasks are different. 
Query-focused summarization tries to find a set of sentences or passages containing frequent and close occurrences of query tokens, while GenEx aims to describe what such sentences convey about the query, explicitly avoiding the query words themselves.
While previous work on snippet generation tries to select or generate a few sentences or even a paragraph, explaining the documents' relevance to a query in a few words is a challenging task.

\subsection{Explainable Search and Recommendation}

How to define and evaluate 
interpretability and explainability of machine learning models are discussed in several studies~\cite{doshi2017towards,Guidotti:2018:SME:3271482.3236009,Lipton:2018:MMI:3236386.3241340,MILLER20191,AAAI1816982}. But, application of machine learning techniques to different tasks can impose task-specific requirements on definition and evaluation of suitable explanations. 
Explaining search results has been briefly  studied in  recent years. 
Describing relationships between entities in queries is considered an explanation of search results and is generated based on incomplete descriptions of relationships in knowledge graph~\cite{voskarides-etal-2015-learning,10.1007/978-3-319-56608-5_25}. As users' information needs are very diverse, search result explanations requires description of much more relevance factors than relationships between entities, which we aim to extract and describe. 
Some models explain search results by providing a set of keywords (with their estimated weights) for each document by training another ranker to simulate the scores of a black-box ranker~\cite{singh2018interpreting,Verma:2019:LLI:3331184.3331377,singh2018posthoc,Singh:2019:EES:3289600.3290620,Roy:2019:ILP:3357384.3357859}. 
These models mainly explain search results following the posthoc explanation method for classifiers - LIME~\cite{Ribeiro:2016:WIT:2939672.2939778}. 
Specifically, Singh and Anand~\cite{singh2018posthoc} interpret a base ranker by training a second tree-based learning-to-rank model with an interpretable subset of content-based ranking features, such as frequency and \texttt{TF-IDF}. Training data for the second ranker is generated from the outputs of the base ranker. 
\citet{10.1145/3397271.3401286} also use basic retrieval heuristics (frequency of a term in a document, frequency of a term in a collection,  and length of a document) as explanation features.
While these interpretable features seem to be useful for system engineers, how to use them to provide explanations for users is unexplored.
 

\citet{Singh:2019:EES:3289600.3290620} uses LIME to explain the output of a ranker, which is based on perturbing the instance to be explained. 
The authors cast the ranking task as a classification task and obtain binary labels of relevant or non-relevant for perturbed documents based on three ways: top-k binary, score-based, and rank-based.
These perturbed instances with their binary labels are then fed to the LIME explainability model and visualized as in the original, using bar-charts to show  word contributions to the model's decision.

\citet{10.1145/3331184.3331312} explore a model-introspective explainability method for neural ranking models. They use the DeepSHAP~\cite{10.5555/3295222.3295230} model to generate explanations and defined five different reference to generate explanations: 1)~document only containing \textbf{OOV} words, 2)~document built by sampling words with low \textbf{IDF} values, 3)~document consisting of words
with low \textbf{query-likelihood} scores, 4)~document sampled from the \textbf{collection} that is not available in the top-1000 ranked list, and 5)~document that is sampled from the \textbf{bottom} of the top-1000 documents retrieved. 
They found that DeepSHAP’s explanations highly depends on a reference
input that needs to be further investigated.
The authors also compared DeepSHAP's explanations with those generated by EXS~\cite{Singh:2019:EES:3289600.3290620} based on LIME and found that they are significantly different. 
They note that this difference  by the two explanation models is concerning, especially in the absence of gold explanations. 
\citet{10.1145/3331184.3331377} propose a model-agnostic  approach based on a weighted squared loss to explain rankers as well as three sampling approaches to perturb the document in the instance to be explained: 
uniform, biased, and masked sampling. Explanation features in their work consist of words. 

Recently, \citet{singh2020valid} propose a   local model-agnostic method for explaining  learning-to-rank models. They define  interpretability features based on IR heuristics and propose two metrics, \textit{validity} and \textit{completeness}, to generate explanations.
They propose a greedy approach to find a subset of the  features such that there is a high correlation between the rankings produced by the selected features and the original black-box model, i.e., high validity. They try to jointly maximize completeness, which is defined as the negative of the correlation between non-explanation features and the original ranking. Correlations between ranked lists are measured using the Kendall's Tau.

Helping users interpret search results can be  different  than, and thus not possible through, presenting (partial) information about how search engines work, such as providing (interpretable) ranking features to users. 
In a  recent study, Thomas et al.~\cite{10.1145/3371390} investigated how users perceive rankings provides by a search engine, with the goal of finding out what forms of explanations may help users. 
 The authors identified six core concepts that used in ranking at Web scale such as \emph{relevance}  and \emph{diversity}. Participants are asked about why each result is chosen. 
 Although \emph{diversity} had been identified as important ranking factor before collecting user responses, less than 1\% of users found that search results are presented because of diversification to cover different intents and facets of queries. In more details, \emph{diversity} has the lowest mentions in the collected responses. 
They mainly performed a set of user studies and online surveys to better understand the mental model of users while using the web search engines. In this work, we propose a model for explaining to what aspect (or facet) of the query, the document is relevant. 
We believe adding our explanations to SERP could address the issue found by \citet{10.1145/3371390} about query intents and facets. 




Explainable recommendation has recently attracted considerable attention~\cite{zhang2018explainable}. For instance, \citet{Ai:2020} recently proposed a model based on dynamic relation embedding to produce explanation for recommendation in the context of e-commerce.
Content-based models for explanation of recommender systems are closer to explanation of search results than those of collaborative filtering, yet structures in items and user-item interactions are not available in Web search.


\subsection{Query Aspects}
Mining query aspects to diversify search results~\cite{Santos:2015:SRD:2802186.2802187} is also related to our work. Most approaches in this category are based on query reformulations found in a query log~\cite{10.1145/1148170.1148320,10.14778/1988776.1988781} or existing taxonomy or knowledge bases such as Open Directory Project~\cite{10.1145/1498759.1498766}. 
Other than these sources, there are some work that extract query facets from the search results. For example,
\citet{10.1007/s10791-013-9221-8} cluster the phrases from top retrieved documents to extract query aspects.
\citet{Kong:2013:EQF:2484028.2484097,10.1145/2661829.2661964} use pattern-based semantic class extraction, such as ``NP such as NP, NP, ..., and NP,   to obtain a list of candidate query facets. Their proposed models based on the directed graphical model or clustering  then filters out noisy candidates. 
Later, \citet{10.1145/2766462.2767757} extend their model to utilize pre-search context in prediction of query intents. 
QDMiner~\cite{7236912} also extract query facets from top search results by using predefined patterns to obtain candidates from free texts. 
In another line,  
\citet{Ruotsalo:2018:IIM:3211967.3231593} propose to model query intent by incorporating feedback from users.  
Although expected explanations to be generated can reveal query aspects, the defined task is different than existing models for mining query aspects as we are interested in explaining the relevant part of a single document's content to a given query without using  other sources of information.  In addition, existing models for extraction of query aspects from top search results are based on predefined patterns, thus the aspects are generated in the extractive setting, while our model generates abstractive explanation for a given query-document pair. 

\citet{10.1145/2348283.2348298} propose AspecTiles to present  the  degree of relevance of a document to each query aspect where the query aspects are given. Our focus in this work is not about what is a good way to present the generated explanations to users, but about how to generate high-quality explanations. Having explanations,  AspecTiles can be one way to present them to users. 
\section{Usability  Study}
\label{sec:usability}

\begin{figure}[t]
\begin{center}
\centerline{\includegraphics[scale=0.6]{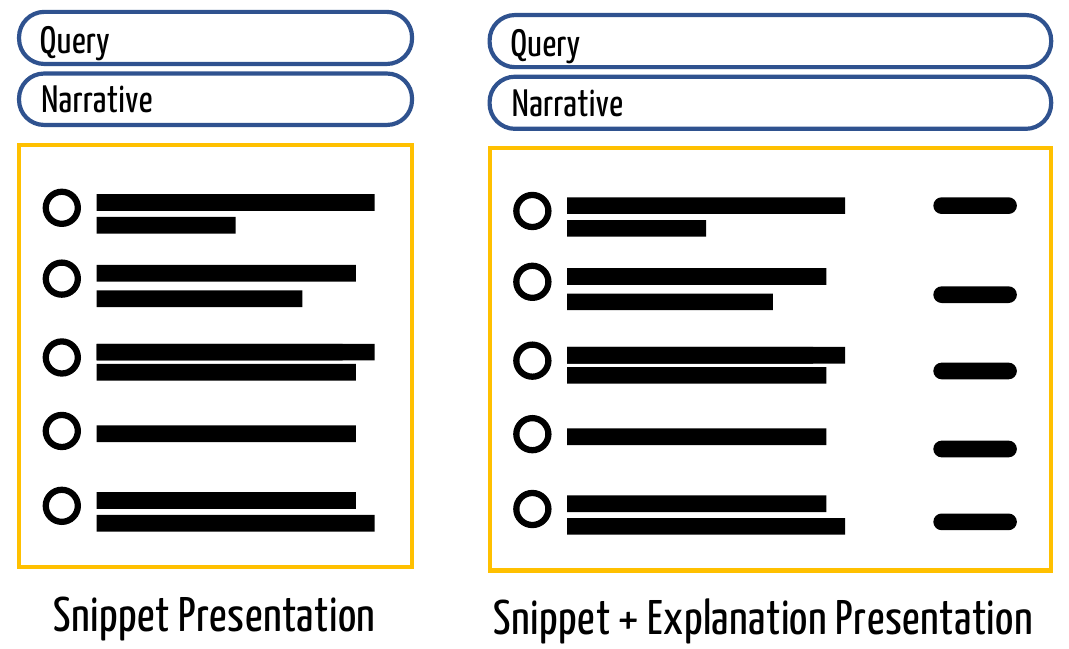}}
\caption{Different schemes of search result presentation for the user study on usability of explanations.}
\label{fig:annotation-ui}
\end{center}
\end{figure}

We start with a user study\footnote{\emph{Location omitted for review} Institutional Review Board  number \emph{omitted-for-review}.} to investigate whether \textsc{GenEx}-style explanations can actually improve the effectiveness and efficiency of search, as two main goals of explanations proposed by~\citet{Tintarev2015}. 
More specifically, we investigate whether providing the explanation to users helps them predict the relevance status of documents faster and/or more accurately. 
We developed two schemes for presentation of search results, illustrated in Figure~\ref{fig:annotation-ui}:
in one scheme (on the left), document are represented by their snippets alone; in the other, documents are presented using the same snippets but also an explanation in a column on the right side of the page. 

We simulate the output of a search engine and creation of explanations for this study. We randomly selected 40 articles from Wikipedia that have at least five sections with headers other than ``stop headers'' such as \emph{references} or \emph{see also}.
Each section of that article is treated as an ``aspect'' of the query that a user might be interested in. We manually developed TREC-style narrative descriptions of the aspects to reduce ambiguity. For four selected articles, the narratives were sufficiently difficult to construct that we discarded the articles. 
The snippets were obtained using the Indri toolkit\footnote{\url{https://www.lemurproject.org/indri.php}}. 

For each query (article title), we created a document per section where the section was deemed relevant to its heading (aspect). By construction, each query had at least five aspects: in the end we had 36 queries leading to 240 unique query-aspect pairs each of which had its section content as a single relevant document. (We will use Wikipedia similarly later in Section~\ref{sec:exp}.)

Each search result in this study was associated with a query-aspect pair. We select the aspect's relevant document and four other documents from the same article so exactly one of the five is relevant. The ordering is random for each pair but is the same for both conditions (with and without the explanation).

We carried out the study on Amazon Mechanical Turk using 
master workers in the United States who had a high  task approval rate. Subjects were provided with search results using one of these presentation schemes and asked to select the document that was relevant to the user's intent (as described by the aspect narrative). A worker was presented with four search results in sequence but could opt to do additional sets. A set comprised four distinct queries and either all included or all excluded explanations. We captured the selected documents  as well as workers' response time. If the mouse was idle for 2 minutes, we assume the worker is not active and reset the timer. Each presentation of the result list of a query-aspect pair is judged by three different workers.

By construction, every query-aspect pair has a relevant document. Users who selected the wrong document more than twice in a set were deemed to have failed -- possibly because they were randomly clicking and not attending to the task -- and the set was rejected (and put back in the pool for annotation). However, if workers felt that there was not a relevant document, we required them to describe why they felt that way. If their reasoning was solid, we accepted the set that would otherwise have been rejected. In the end, 240 query-aspect pairs were presented in two styles (with or without explanation) and annotated by three distinct workers, for a total of 1,440 accepted judgments. A total of 51 unique workers participated.

\begin{table}
	\caption{Usability study results.}
	\label{table:usability-user-study-results}
	\centering 
	\begin{tabular}{ l  l c c c } \hline
	\multirow{2}{*}{Style} & \multirow{2}{*}{$\mathcal{K}_f$} & Correct  & Majority & Avg. res.     \\ 
	&  & relevant & relevant & time (s)  \\ \hline
	Snippet only    & 0.67 & 66\%  & 168 (70\%) & 35.7  \\ \hline
    Snippet + Expl. & 0.92 & 91\%  & 224 (93\%) & 23.1    \\ \hline
	\end{tabular}
\end{table}

Recall that our question is whether the explanation helped users identify relevant documents faster and/or more accurately. Table~\ref{table:usability-user-study-results} summarizes the results of this study. To consider accuracy, we first look at the agreement among the three annotators, measured by Fleiss’ Kappa  ($\mathcal{K}_f$)~\cite{Schouten1986}. We find that agreement is substantially greater with the explanation: $\mathcal{K}_f$ increases by 37\% from 0.67 (substantial agreement) to 0.92 (almost perfect). The fraction of all judgments that are correct increases by 38\% with the explanation and the proportion of the 240 instances where the majority judgment is correct show a 33\% climb from 168 to 224. We conclude that the explanation presented greatly increased the consistency and accuracy of identifying relevant documents. 

We also compare the average response time for query-aspect pairs. 
The average time for selecting the correct relevant document decreases from 32.4 to 25.4 seconds when explanations are also provided, which is a 22\% decrease.
To be sure we were ignoring times when the worker was perhaps unfaithful to the task (so rapidly clicking), we only include cases where the majority of workers selected the relevant document in \emph{both} presentation schemes. There was one case that the majority voted on non-relevant documents when explanations are provided, but voted on the relevant document given just the snippets. The average response time is thus evaluated over 167 query-aspect pairs. Each query-aspect pair thus had 2 or 3 correct responses. We report the macro-average of response times for correct responses in Table~\ref{table:usability-user-study-results}, showing a 35\% decrease when explanations are also provided for search results.

We highlight that this study used manually generated search results and relevance judgments. Nonetheless, the results strongly suggest that adding explanations to snippets can greatly improve both the accuracy and speed of judging documents for relevance. Buoyed by those results, we next propose \textsc{GenEx}, an approach for creating these explanations.

\section{Generating Terse Explanations} 
\label{sec:model}

Neural approaches to explanation generation conceptualize the task as a sequence transduction problem. A major approach to this problem is based on the encoder-decoder architecture, where an encoder processes the input tokens and a decoder generates explanation tokens, autoregressively. 
In the defined task of explaining documents' relevance to a search query, the encoder is expected to learn a contextual representation of the document, capturing the query related parts of the document. The decoder, on the other hand, should generate a text that explains how the document is relevant to the query. In this section, we introduce \textbf{GenEx}, our approach for \textbf{gen}erating documents' relevance \textbf{ex}planations.

\subsection{Problem Formulation}
Given a query $\mathbf{q}$ and a document $\mathbf{d}$, the goal is to learn an abstractive explanation  model $\mathbf{e} = \mathcal{F}(\mathbf{q}, \mathbf{d})$ to generate a text sequence $\mathbf{e}$ that explains how the given document $\mathbf{d}$ is relevant to the query $\mathbf{q}$. The explanation generated by the model may include tokens that do not actually occur in the document.
A training instance for this task is thus denoted as the triplet $(\mathbf{q}, \mathbf{d}, \mathbf{e})$.

\begin{figure}[t]
\begin{center}
\centerline{\includegraphics[width=0.7\columnwidth]{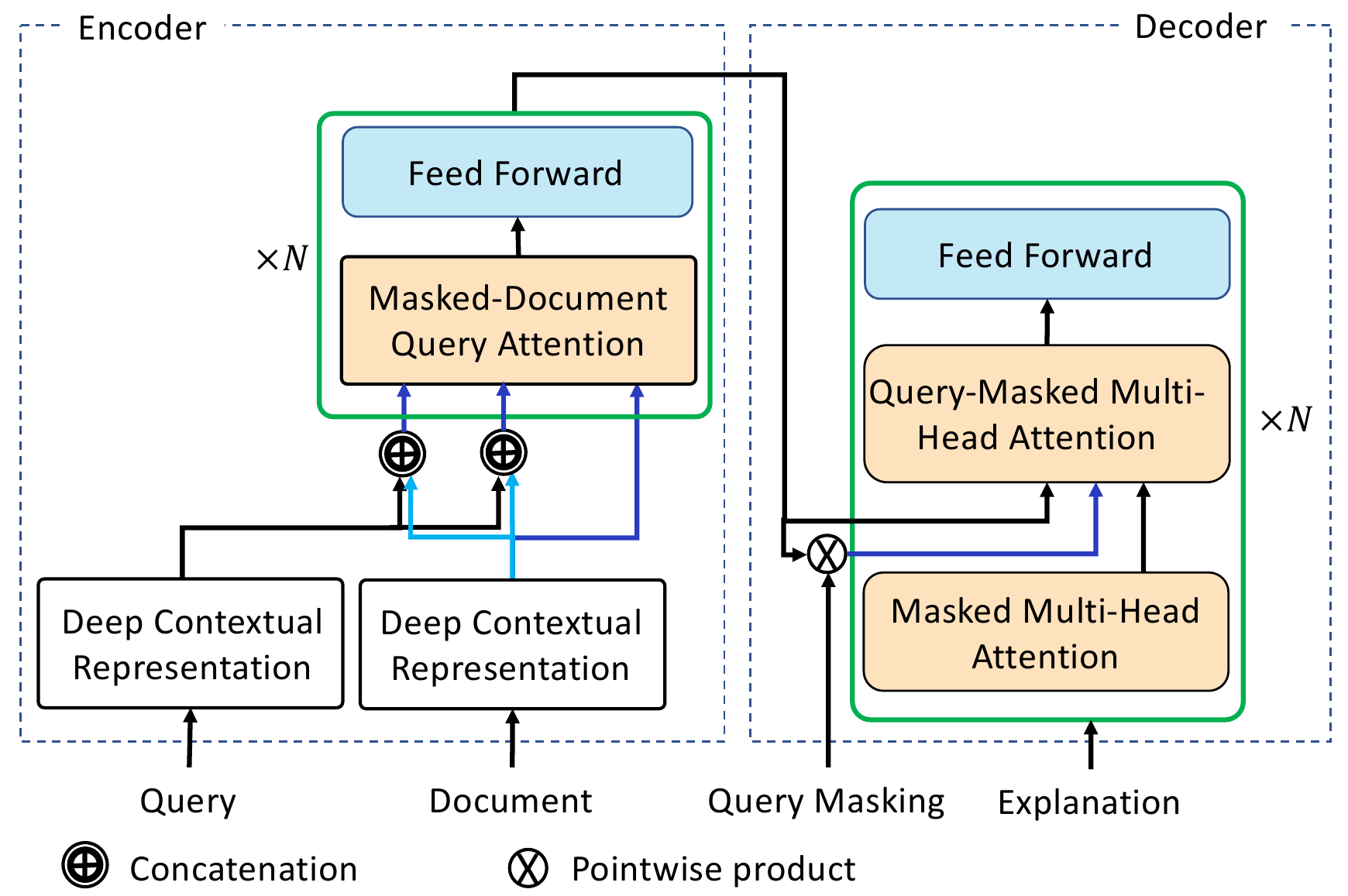}}
\caption{GenEx Architecture (residual connection and layer normalization for each sub-layer in green boxes have been omitted to enhance the clarity of the figure).}
\label{fig:arch}
\end{center}
\end{figure}

\subsection{Input Representation}

We first tokenize queries, documents, and target explanations  using subword tokenization following Wu et al.~\cite{wu2016google} and encode tokens  with a pre-trained vocabulary embeddings~\cite{devlin-etal-2019-bert}.

We then add \textbf{positional encodings} to the \textbf{token embedding} using the sine function following the Transformer model~\cite{Vaswani:2017:AYN:3295222.3295349} to capture the relative positions of tokens in input and target sequences given to the encoder and decoder, respectively.

Finally, as the input to the explanation task consists of  query and document parts, 
we add a special token at the end of each part of the input. We also construct   a \textbf{segment embedding}   in two ways as the input to our explanation models:~(1)~segment embeddings indicating whether a token belongs to the query or to the document, and (2)~embeddings indicating whether a token occurs in the query or not. 
The choice of segment embedding is based on the model architecture and will be discussed later.

\subsection{Contextual Encoding of Query and Document}
A straightforward solution to handle the two-input characteristic of the explanation task
is to concatenate  query  and  document tokens separated using a special token, and then feed the obtained vector to the encoder of a sequence-to-sequence model, such as the Transformer encoder~\cite{Vaswani:2017:AYN:3295222.3295349}.
The input sequence to the model is thus as follows:
\begin{equation}
    (q^1, \dots, q^m, \sigma, d^1, \dots, d^n),
    \label{baseline_input}
\end{equation} 
where $\sigma$ is a special separation token, similar to the input of several Bert based models for different tasks such passage/document ranking~\cite{nogueira2019passage} or question answering.
We observed, however, that concatenation of short keyword queries with long documents does not lead to sufficient attention weights from document tokens to the query tokens. This is while learning a query-focused  representation of documents is crucial for generating explanations.

To learn proper attention from  document tokens to query tokens, we first 
build two  input sequences for a given query and document.  
These inputs are then separately fed  to the Transformer-based encoder model with shared weights. The self-attention mechanism  in the Transformer leads to contextual representations of query and document tokens.
The query and document representations obtained by this encoding component are denoted as $\boldsymbol{z}_q$ and $\boldsymbol{z}_d$, respectively.
This architecture also allows pre-computation of document representations similar to the recent Bert-based rankers~\cite{10.1145/3397271.3401075,10.1145/3397271.3401093}.

\subsection{Query Attention Layer} 
To obtain query-focused representations of documents, we propose another encoder model on top of the learned contextual representations of document tokens. The architecture is shown in~\figurename~\ref{fig:arch}. 
This second encoder model is based on the Transformer encoder where  self-attention layers are replaced with masked-document query attention layers. 
This layer consists of three parameter matrices $W_Q$, $W_K$, and $W_V$. The query attention layer computes the following attention matrices based on inputs different from the self-attention mechanism: 
\begin{align}
    Q &= \boldsymbol{z}_d \times W_Q, \\
    K &= \big(\boldsymbol{z}_q \| \boldsymbol{z}_d \big)  \times W_K, \\
    V &= \big(\boldsymbol{z}_q \| \boldsymbol{z}_d \big) \times W_V, 
\end{align}
where $\|$ donates concatenation of representations build from previous encoder layers. 
The representation of each document token is then updated using the scaled dot-product attention function of the Transformer as below:
\begin{equation}
    \mathrm{Attention} (Q, K, V) =  \mathrm{softmax}\big(\frac{QK^T}{\sqrt{k}}\big)V,
    \label{eq:self.attention}
\end{equation}
where $k$ is the dimension of the keys $K$. 
However, we mask the scaled dot product  in a way to prevent a document token from attending to other document tokens. 
Therefore, the output representation of a document token is computed as a weighted sum of the attention values corresponding to itself and query tokens. 
Lastly, representations are obtained by concatenation of outputs from multiple attention functions, each projecting its inputs to a different subspace. 
The input and output dimensionality of query attention layers are the same and a stack of $N$ query attention layers is used in the second encoder component. 
The representations of document tokens from the second encoder component constitute the input of the decoder.

\subsection{Query-Masked Decoding}
Documents on top of a search engine result page often contain  query tokens with high frequencies~\cite{manning2008introduction}. In addition, our encoder consists of  masked-document query attention layers, which highlight the query related part of the document. 
Therefore, it is likely that the decoder generates  query tokens as the explanation of a document's relevance. However, this is not desired; the model should generate explanations that provide more information other than the query itself, otherwise the generated explanation is useless. Therefore, we generate  explanations by extending the Transformers Decoder architecture~\cite{Vaswani:2017:AYN:3295222.3295349} using a query masking mechanism to reduce the probability of generating query tokens by the decoder.
To achieve this, we use a \emph{masked} multi-headed attention for the encoder-decoder attention layer in which the query tokens in both query and document are masked. This makes every decoder position to attend over only non-query tokens in the input sequence.

Note that query tokens are not replaced with a special mask token, because these tokens are central in building the representation of documents. Instead, the representations of query tokens are masked during the decoding process to allow the model to describe the query-related information that the document provides.

\section{Experiments}
\label{sec:exp}

\begin{table}
	\caption{\texttt{Wiki} dataset statistics, where each sample consists of a query-document-explanation triple, and average values of length  are calculated on the specified element of samples. 
	}
	\label{table:stat}
	\centering 
	\begin{tabular}{ l c c  c } \hline
& Train & Dev & Test  \\ \hline
number of samples &  6,386,916 & 336,425 & 10,000 \\ \hline
average length of queries  & 4.2 & 4.2 & 4.2\\ \hline
average length of documents  & 231.8 & 231.8 & 233.2 \\ \hline
average length of explanations & 2.4 & 2.5 & 2.5\\\hline
	\end{tabular}
\end{table}

\subsection{Datasets}

We constructed \textbf{Wiki training data} using 
\emph{Wikipedia articles}.  In this dataset, each section of a Wikipedia article is treated as a document. The section documents extracted from a Wikipedia article are all relevant to  the query built from the article's title, with their section headers as their explanation labels. 
For the work discussed here, we use a March 2019 dump of English Wikipedia. 
Only Wikipedia articles are used for building the dataset, and Wikipedia pages such as disambiguation and redirect pages as well as  administration pages such as talk, user, and maintenance pages are removed. 
Articles with URL links as their titles are also removed.
We extract all text from each Wikipedia article, filtering out unwanted data such as HTML markup, using WikiExtractor\footnote{{\url{https://github.com/attardi/wikiextractor}}}
and then discard any article that has fewer than 500 characters of text. A section is not included unless it contains at least 20 tokens. 
Sections with highly-frequent headers such as ``references'' and ``see also'' are removed. 
The obtained samples are randomly divided into train, validation, and test samples referred to as the \texttt{Wiki} test set. 
Statistics of Wiki dataset are reported in Table~\ref{table:stat}. The reported length values are based on tokens obtained by using sub-word tokenization. 
Explanations having a token not occurred in the query or document are counted as abstractive explanations. Almost 98\% of explanations have a token not occurred in the input sequence.

We also built \textbf{Anchor training data} to improve the generalizability of  learned explanation models. Ideally, we could have used Web search logs for mining different facets of  a head query, but they are not publicly available due to concerns about user  privacy. 
However, it has been shown that \emph{anchor text}  can effectively simulate real user logs for query reformulation techniques~\cite{Dang:2010:QRU:1718487.1718493}. Based on that work, we approximate query facets using widely available anchor text data, where different anchor texts linking to a particular page approximate query facets relevant to that page. 
For this purpose, we follow the \emph{subtopic clarification by keyword} observation through  user log analysis~\cite{Hu:2012:MQS:2348283.2348327}. We used the external version of the \emph{anchor text for ClueWeb09} dataset\footnote{\url{http://lemurproject.org/clueweb09/anchortext-querylog/}}
to build training data for the explanation task. All anchor texts to one page that start with the same prefix words are grouped together. The common prefix is then considered as a query, and each suffix is considered as a facet of the query. The linked page is considered to be a relevant document to the query. We performed a number of post-processing steps on the extracted query facets. For examples, facets containing words such as ``homepage'' or ``website'' are disregarded. Stopwords such as ``of'' or ``and'' are removed from the beginning of query facets.

\textbf{Clue-Res test data} is generated from the ClueWeb09 category B dataset\footnote{\url{http://lemurproject.org/clueweb09/}} 
which is a standard TREC collection and has been used in the TREC Web track for several years~\cite{collins-thompson2014trec,collins-thompson2015overview}.
Topics from TREC Web track 2009 to 2012 are used to build test samples. 
We use the title of a TREC topic as the query, and subtopics of relevant documents in judgments as the set of explanations. 
Some topics have multiple subtopics. Navigational subtopics as well as the first subtopic of queries are discarded since first subtopics mainly provide general description of queries and are not focused around a query facet. 
Subtopic descriptions are manually rephrased by removing phrases such as ``I'd like to find information about''. 
This test set contains 543 samples.

We built \textbf{Passex test data} using the passage ranking dataset of TREC 2019's {Deep Learning} Track\footnote{\url{https://microsoft.github.io/TREC-2019-Deep-Learning/}}.
The TREC passage ranking dataset is built based on the MS MARCO dataset where passages retrieved with respect  to the test questions are judged for relevance in much more details. Questions having more than 30 relevant passages in the dataset are chosen for building our evaluation samples. Questions are then manually rephrased as short keyword queries and explanations. 
Some questions whose conversion to query and explanation were not straightforward are ignored in this step. Also relevant passages with less than 100 tokens are removed. In the end, there are 27 unique queries and 188 evaluation samples in the built Passex dataset. 

Note that target outputs in the built test samples are not precise  explanations. These test sets are constructed to have approximate out-of-domain samples to guide the development of a generalized explanation model. In the current study, we focus on explaining why a document is relevant to a query by providing the query aspect that the document covers as supportive evidence. Explanation of non-relevant documents   to a query is left for future work.

\textbf{Pre-processing steps.}
Document texts, queries, and ground truth explanations  are lowercased, and encoded using sub-word tokenization following Wu et al.~\citeyearpar{wu2016google} with Bert vocabulary\footnote{\url{https://storage.googleapis.com/bert_models/2018_10_18/uncased_L-12_H-768_A-12.zip}} 
of size 30,522. 
We consider only documents that have more than 20 tokens, and truncate long documents. 
Documents of the Wikipedia dataset  are truncated if required by keeping the first $L$ tokens.
We do not truncate the output sequence, but we ignore training samples whose explanations (sub-headings) are longer than 15 tokens. 
Documents chosen from the ClueWeb dataset for building anchor dataset are extractively summarized as below.

\textbf{Extractive summarization of  long documents.}
Documents are segmented into sentences using the NLTK toolkit\footnote{\url{https://www.nltk.org/}}. 
Sentences with exact occurrences of query terms where document and query terms are stemmed using the Porter stemmer, are chosen to be in the query-biased extractive summary of the document. 
To capture all query-related information of the document, we also compare the Word2vec pre-trained embeddings of all pairs of  query-document terms. A sentence whose most similar term to a query term has a  score higher than a defined threshold of 0.8 is retained for the summarized version of the document. If the length of selected sentences is less than $L$, then additional
sentences with the highest contextual similarity to the query are added in the order of their occurrence until no other sentences can be added to the summary while keeping the length lower than or equal to the length cap $L$.

\subsection{Evaluation Metrics}
Following the text summarization and machine translation communities, we use 
BLEU~\cite{Papineni:2002:BMA:1073083.1073135} and  different variations of the
ROUGE metric~\cite{lin-2004-rouge} such as   ROUGE-1 (unigrams),  ROUGE-2  (bigrams),  and  ROUGE-L (longest-common substring)\footnote{\url{https://github.com/google-research/google-research/tree/master/rouge}}
for evaluation of generated explanations: the more words and n-grams that are in common between the  predicted output and the target subtopics, the more likely the explanation is to be good.
We also use BERTScore~\cite{Zhang*2020BERTScore:} to semantically compare  generated explanations with reference explanations. 
BERTScore uses pre-trained contextual embeddings from BERT or its variants to compute cosine similarity between tokens in candidate and reference text segments. 
Zhang et al.~\cite{Zhang*2020BERTScore:} showed that BERTScore better correlates with human judgements and thus is a stronger metric for comparing text generation models. 
Two-tailed paired $t$-test is used to test whether the differences between performance of models are statistically significant. 
Content-based explanation of a single document with respect to a query does not produce any rankings,  therefore metrics for evaluation of rankings such as Mean Average Precision are not suitable to measure the quality of explanations.

\subsection{Baseline Models}
To the best of our knowledge, this is the first study to explain search results from a black-box ranker by generating abstractive and concise explanations beyond term matching. 
The unique characteristics of the defined problem make models for related task not suitable where the differences are discussed in Section~\ref{sec:rel}.
The first baseline is to use TextRank algorithm~\cite{mihalcea-tarau-2004-textrank}  to extract document keywords. 
The TextRank algorithm represents a document as a graph of terms linked by co-occurrence relation. 
Two terms are connected if they co-occur within a window of fixed size, set to 10 terms in our experiments. 
Then, the PageRank algorithm is applied to the graph to rank terms. The top ranked terms of a document are identified as its keywords. 
For the defined explanation problem, we are interested in describing a document with respect to a query. To accommodate this setting, we also tried the TextRank algorithm by using  topic-sensitive PageRank~\cite{10.1145/511446.511513} to rank graph vertices. We refer to this modified version as \emph{topic-sensitive TextRank}.
We also compare with KeyBERT~\cite{grootendorst2020keybert} which uses BERT to extract keyphrases from a document.
This baseline demonstrates the necessity of attending to queries for explanation of document relevance, therefore models for keyword or keyphrase extraction from documents cannot fulfill the task of relevance explanation with noun phrases. 
Due to the lack of  training data for query-focused keyword extraction, we chose the unsupervised and widely used TextRank model as well as recent BERT-base model KeyBERT.

The next group of baselines is  based on using local model-agnostic explanation methods to describe the relevance of a document with respect to a query. 
For this group, we use LIME~\cite{Ribeiro:2016:WIT:2939672.2939778} and Sensitivity~\cite{10.1007/978-3-319-10590-1_53} that explain the prediction of a black-box classifier for a given input sample.
For the purpose of our task, explanation features are defined as document tokens~\cite{Ribeiro:2016:WIT:2939672.2939778,Singh:2019:EES:3289600.3290620}.
Specifically, the LIME method generates a number of  perturbed samples of the input  and learns a linear model, based on the classifier's predictions for perturbed samples.
We trained a linear SVM model as the explanation model in our experiments. 
Features with large coefficients in the learned linear model are considered as explanation. 
The Sensitivity method estimates the score of a token by measuring the change in the predicted relevance probabilities of perturbed samples that do not include that token~\cite{10.1007/978-3-319-10590-1_53}.
The higher the contribution of a token in the relevance probability is, the more important the token is in understanding the relevance of the document with respect to the query.

We used two different types of rankers as the black-box model whose predictions are to be explained by LIME and sensitivity. These rankers are used to get the relevance probabilities of perturbed documents with respect to the query.
The first ranker for this purpose is the classic TF.IDF ranker. For this ranker, documents are represented as bag of words and perturbed samples are obtained by randomly changing one dimension of document vectors to zero. 
We also used the BERT-based ranker~\cite{nogueira2019passage} since it has been shown to achieve the current state-of-the-art performance~\cite{craswell2020overview,gao-etal-2020-modularized}.
In this ranker, document tokens that are semantically similar to query tokens can impact the relevance probability of a document, in contrast to the TF.IDF ranker which is only based on exact term matching between a query and document. 

The input of the BERT-based ranker is built by concatenating the query and document with a separate token. 
The pre-trained BERT-based model is fine-tuned for the ranking task using the training data of the MS MARCO passage-ranking dataset~\cite{nogueira2019passage}. 
As the BERT-based ranker uses sub-word tokens,
the term-level score is obtained by summing the score of each token's subword. 
For BERT-based ranker, a given input sample
 which is a query-document pair in the setting of our task,
is perturbed by  masking a token from a random position in the document.
The final score of each document token is calculated by summing the scores obtained for all occurrences of the token in the document. 
Document tokens are sorted by the obtained scores and the top-ranked tokens are considered as the explanation of the document. The top-ranked tokens are considered to be tokens whose scores are not less than 10\% of the token with the highest score.
We consider the top three tokens if more than 3 are selected by the thresholding function. 
The cut-off value is chosen based on the average length of gold explanations in the test sets. Note that the average length of gold explanations is not used by the GenEx model.

The last baseline is training the original Transformer model where input sequences are constructed by concatenating query and document tokens in the training dataset as Eq.~\eqref{baseline_input}. 
The model input is constructed  by adding  segment embeddings differentiating query and document segments in addition to the separator token $\sigma$ between them.  
These segment embeddings have the value of 0 for query tokens and 1 for all document tokens.
We refer to this model as \texttt{Orig-Trans}.

\begin{table*}
	\caption{Performance of different explanation models. Symbols $^\blacktriangledown$ and  $^\triangledown$ show statistical significant differences with GenEx at levels 0.01 and 0.05, respectively.}
	\label{table:baseline-kw}
	\centering 
	\begin{tabular}{ l c c  c c c c}
	\cline{2-7}
    & \multicolumn{2}{c}{Wiki} & \multicolumn{2}{c}{Clue-Res} & \multicolumn{2}{c}{Passex}  \\ \cline{2-7}
    &  BLEU-1 & R-1  &  BLEU-1 & R-1  &  BLEU-1 & R-1  \\ \hline
    TextRank & 0.0862$^\blacktriangledown$ & 0.1427$^\blacktriangledown$ & 0.0331  & 0.0435 & 0.0319$^\blacktriangledown$ & 0.0880$^\blacktriangledown$ \\ \hline
    TS-TextRank & 0.0736$^\blacktriangledown$ & 0.1169$^\blacktriangledown$ & 0.0313$^\blacktriangledown$ & 0.0369 & 0.0248$^\blacktriangledown$ & 0.0715$^\blacktriangledown$ \\ \hline
    KeyBERT & 0.0567$^\blacktriangledown$ & 0.0666$^\blacktriangledown$ & 0.0437 & 0.0404 & 0.0160$^\blacktriangledown$ & 0.0222$^\blacktriangledown$  \\ \hline
    LIME + TF.IDF & 0.0454$^\blacktriangledown$ & 0.0366$^\blacktriangledown$ & 0.0281$^\blacktriangledown$  & 0.0288$^\blacktriangledown$ &  0.0059$^\blacktriangledown$ & 0.0044$^\blacktriangledown$  \\ \hline
    LIME + BERT & 0.0104$^\blacktriangledown$ & 0.0085$^\blacktriangledown$ & 0.0178$^\blacktriangledown$ & 0.0257$^\blacktriangledown$ & 0.0089$^\blacktriangledown$ & 0.0418$^\blacktriangledown$ \\ \hline
    Sensitivity + BERT & 0.0062$^\blacktriangledown$ & 0.0046$^\blacktriangledown$ & 0.0330$^\triangledown$ & 0.0398$^\blacktriangledown$ & 0.0019$^\blacktriangledown$ & 0.0018$^\blacktriangledown$ \\ \hline
    GenEx &  \textbf{0.2313}$\;$ & \textbf{0.3582}$\;$ & \textbf{0.0520}$\;$ & \textbf{0.0617}$\;$ & \textbf{0.1264}$\;$  & \textbf{0.1179}$\;$ \\ \hline
	\end{tabular}
\end{table*}

\begin{table*}
	\caption{Performance of generative explanation models based on R-1 F-measure, R-2 F-measure, and R-L F-measure metrics. 
Symbols $^\blacktriangle$ and  $^\triangle$ show statistical significant differences with Orig-Trans at levels 0.01 and 0.05, respectively.
	}
	\label{table:perf-ROUGE}
	\centering 
	\begin{tabular}{ l c c  c}
	\hline
& \multicolumn{3}{c}{Wiki} \\ \cline{2-4}
 &  R-1 & R-2 & R-L \\ \hline
 
 Orig-Trans &  0.3180 & 0.0940 & 0.3173  \\ 
 
Seg-q-toks &  \textbf{0.4103}$^\blacktriangle$ & \textbf{0.1251}$^\blacktriangle$ & \textbf{0.4086}$^\blacktriangle$   \\ 

Sep-q-doc &  0.4056$^\blacktriangle$ & 0.1247$^\blacktriangle$  &  0.4044$^\blacktriangle$  \\ 

GenEx & 0.3582$^\blacktriangle$ & 0.0868 & 0.3575 \\
\hline
	\end{tabular}

		\begin{tabular}{ l c c  c }
		\hline
&  \multicolumn{3}{c}{Clue-Res}   \\ \cline{2-4}
 &  R-1 & R-2 & R-L   \\ \hline
 
 Orig-Trans &    0.0421 &  0.0005 & 0.0418 \\ 
 
Seg-q-toks  & 0.0515 & 0.0031 & 0.0510  \\ 

Sep-q-doc &  0.0572 & 0.0043 & 0.0562  \\ 

GenEx &  \textbf{0.0617}$^\triangle$ & \textbf{0.0055} & \textbf{0.0614}$^\triangle$  \\
\hline
	\end{tabular}
	\begin{tabular}{ l c c  c }
    \hline
&   \multicolumn{3}{c}{Passex}  \\ \cline{2-4}
 &  R-1 & R-2 & R-L \\ \hline
 
 Orig-Trans  & 0.0780 &  0.0319 & 0.0752\\ 
Seg-q-toks  & 0.0975 & 0.0372 & 0.0993 \\ 
Sep-q-doc & \textbf{0.1330}$^\triangle$ & {0.0426} & \textbf{0.1339} \\ 
GenEx &  0.1179$^\triangle$ &\textbf{ 0.0479} & 0.1179$^\triangle$ \\
\hline
	\end{tabular}
\end{table*}

\begin{table*}
	\caption{Performance of generative explanation models based on BLEU metrics. Symbols $^\blacktriangle$ and  $^\triangle$ show statistical significant differences with Orig-Trans at levels 0.01 and 0.05, respectively. 
	}
	\label{table:perf-BLEU}
	\centering 
	\begin{tabular}{ l c c  c c c c}
	\cline{2-7}
& \multicolumn{2}{c}{Wiki} & \multicolumn{2}{c}{Clue-Res} & \multicolumn{2}{c}{Passex}  \\ \cline{2-7}
 &  BLEU-1 & BLEU-2  &  BLEU-1 & BLEU-2  &  BLEU-1 & BLEU-2  \\ \hline
 
Orig-Trans &  0.2269$\;$ & 0.1357 & 0.0301 & 0.0056 & 0.0755 & 0.0468  \\ 
Seg-q-toks & 0.2918$^\blacktriangle$ & 0.1791$^\blacktriangle$ &  0.0385 & 0.0108  & 0.1038 &  0.0595 \\
Sep-q-doc & \textbf{0.2979}$^\blacktriangle$ & \textbf{0.1845}$^\blacktriangle$ & 0.0473$^\triangle$& \textbf{0.0156}$^\blacktriangle$  & 0.1212$^\blacktriangle$ &   0.0651\\
GenEx & 0.2313$^\blacktriangle$ & 0.1236 & \textbf{0.0520}$^\blacktriangle$ & 0.0099$^\blacktriangle$ & \textbf{0.1264}$^\blacktriangle$ & \textbf{0.0718}\\
\hline

	\end{tabular}
\end{table*}

\textbf{Ablation study.}
There are two main architectural choices in the proposed explanation model; query attention layer, and query-masking during the decoding. To show the effectiveness of each choice, we compare our final model with the following variants, gradually constructed on top of the \texttt{Orig-Trans} model. 
The first variant uses separate encoders for queries and documents with an additional encoder for documents which  consists of  query attention layers. However, query tokens are not masked in the decoder. We refer to this model as \texttt{Sep-q-doc}.
We also test another variant by segment embeddings that differentiate query tokens from other tokens in the input sequence, having 0 for query tokens in the input sequence and a value of 1 for other tokens. Therefore, query tokens in a document are also masked during the decoding process, and only the final hidden vectors corresponding to non-query tokens in a document are fed into the decoder. This is similar to the decoder input of the GenEx model. This variant is referred to as \texttt{Seg-q-toks} model.

\begin{table}
	\caption{Performance of generative explanation models based on F1-score of BERTScore. Statistical significant differences between the GenEx and Orig-Trans at the levels of 0.01 and 0.1 are shown with $^\blacktriangledown$ and $^\triangledown$, respectively.
	}
	\label{table:perf-Bertscore}
	\centering 
	\begin{tabular}{ l c c c} 
	\hline
& Wiki & Clue-Res & Passex \\ \hline 
Orig-Trans &  0.4071$^\blacktriangledown$  & 0.2985$^\blacktriangledown$ & 0.3596$^\triangledown$\\ 
Seg-q-toks & \textbf{0.4796} & 0.3172 & 0.3989 \\
Sep-q-doc & 0.4705 & 0.2898 & 0.3786\\
GenEx & 0.4472 & \textbf{0.3303} & \textbf{0.4430}\\
\hline
	\end{tabular}
\end{table}

\begin{table*}
	\caption{Examples of  explanations generated by GenEx, TextRank, and LIME-TF.IDF.}
	\centering 
	\begin{tabular}{ l l  } \\ \hline
	Query  & diversity \\ \hline
    Doc.   title &  Equality \& Diversity: Athena Report and Action Plan \\ \hline 
    TextRank &  ``oxford",
``athena",
``women" \\ \hline
LIME+TF.IDF & diversity \\ \hline
    GenEx  &  women s career development\\
    \hline 
    \end{tabular}
    \quad
    \begin{tabular}{  l l  } \\ \hline
	 Query & OCD\\ \hline
     Doc. title & OCD  Obsessive-Compulsive Disorder - Mahalo \\ \hline
     TextRank & ``ocd'',	``disorder'', 	``help''\\ \hline
     LIME+TF.IDF & ocd \\ \hline
     GenEx & obsessive \\ 
 \hline
	\end{tabular}
	\label{table:example}
\end{table*}

\begin{table*}
	\caption{Example  documents  from the Clue-Res dataset. 
	Only the beginning  of the first document is  copied here. 
    }
	\label{table:doc_content_examples}
	\centering 
	\begin{tabular}{ l p{12cm}  } \\ \hline
    Doc. 1 & university of oxford athena project 2000 1 action plan encouraging applications from women scientists summary in 1999 2000 the university of oxford applied successfully to the national athena project 1 for funding to assist with a programme of positive action aimed at encouraging applications from women scientists for academic appointments at the university . positive action with the objective of encouraging applications from an under represented sex is defined and authorised by the sex discrimination act 1975 . the university s application was based on an analysis of data from its recruitment monitoring scheme . this consistently demonstrates that women are appointed to academic posts , including those in science , engineering and technology set , at least in proportion to their applications , but that the rate of applications from women is low compared with numbers suggested by the available data to exist in recruitment pools such as contract research staff at oxford and elsewhere and lecturers at other institutions . the acceptance of athena funding commits the university to develop and carry out an action plan based on the experience of its project . this action plan pdf file , 12kb is annexed in tabular form and further information on the oxford athena project is provided below . although the oxford athena project , and therefore this report , deal with the scientific disciplines , there is evidence that women are similarly under represented at oxford in some areas of  
    the humanities and social sciences and , where appropriate , the action plan is intended to cover all disciplines . \\ \hline \hline
    Doc. 2 & obsessive compulsive disorder ocd is an acronym for the mental disorder known specifically as obsessive compulsive disorder . ocd is a type of anxiety disorder . ocd can persist throughout a persons life . the symptoms of ocd can be mild to severe . if severe , they can interfere with a persons ability to function at work , school and home . symptoms ocd involves uncontrollable urges , rituals , or thoughts that cannot be put out of a persons mind . such behavior , may be all consuming , and eventually take over the person s life . a person with ocd feels the need to repeat the same thing over and over to keep bad things from happening . the symptoms of ocd consist of obsession , the constant idea that something bad is going to happen , and compulsion , the constant action to try to prevent the bad things . for example , someone that is worried about germs will wash their hands over and over .  cause the precise cause of ocd is still not known . while some researchers believe it is a chemical imbalance , others see it as a physical condition . \\ \hline

    \hline
	\end{tabular}
\end{table*}

\subsection{Experimental Details}
All baseline based on the Transformer architecture and GenEx are all built and trained using the same settings of hyperparameters and on the same training data to ensure a fair comparison between them.
Hyperparameters are mainly set following the base Transformer model. 
Each encoder/decoder module consists of a stack of 6 identical layers. 
We used 8 attention heads in each layer, each with depth of 96. The input dimension is 768.
Deep contextualized representations of queries and documents can be obtained by fine-tuning BERT. However, our pilot experiments showed that we can obtain reasonable contextual representation using 6-layer Transformer encoder, mainly because these representations will be updated by the next encoder component.  Therefore, considering computational constraint and having large amount of weakly labeled training data, we decided to use Transformer encoder with pre-trained input embeddings from BERT, instead of using the entire pre-trained BERT model. 
Training samples are batched by their total length. Each batch consists of samples with approximately 2048 tokens in total. The input sequence to encoder modules are truncated if required by keeping the first 256 tokens.  
The models are trained for 10 epochs. 
We used the Adam optimizer~\cite{adam-opt} with $\beta_1 = 0.9$, $\beta_2 = 0.98$,  $\epsilon = 10 ^ {-9}$, and a fixed learning rate of 0.00001. Dropout rate of 0.1 is used during the training of models. 
In addition, target outputs are smoothed with a value of 0.1. 
Explanations for test samples are generated by greedy decoding.

 \subsection{Results and Analysis}
 In this section, we provide the evaluation results of the proposed model.
Performance of GenEx and baseline models on Wikipedia test fold (\texttt{Wiki}), \texttt{Clue-Res}, and \texttt{Passex} datasets in terms of  ROUGE and BLEU metrics is shown in Table~\ref{table:baseline-kw}.
As baseline models do not consider the order of words that are selected as explanation, we do not compare them with GenEx based on the performance metrics that depend on higher order n-grams.
The difficulty of the task at hand, the limitations of the noisy automatically-generated training data, and imprecise labels of test data are the main reasons for low performance values. 
GenEx outperforms all baselines over all three test sets where the improvements are mostly substantial and statistically significant. The BLEU improvements of GenEx over the best performing baseline are 168\%, 18\%, and 296\% over Wiki, Clue-Res, and Passex datasets respectively. In terms of the BLEU metric, the best performing baseline over Wiki and Passex is TextRank, and KeyBERT shows the highest performance over the Clue-Res test set. In terms of the ROUGE metric, the best performing baselines over all datasets is TextRank. 
The ROUGE improvements of GenEx over TextRank are 151\%, 41\%, and 34\%.

The first group of baselines is keyword extraction by TextRank and its topic-sensitive variant.
TextRank provides a query-independent explanation of a document, while topic-sensitive TextRank is developed to explain documents with respect to a given query. 
As a small part of a document may be relevant to the query~\cite{TREC} and in such cases we are not interested in the document keywords related to its general topic, the topic-sensitive variant of the TextRank algorithm should intuitively select document words that are more suitable for explanation of document relevance.
The reported results in Table~\ref{table:baseline-kw} show that  these models are among the best performing baselines. 
However, the TextRank algorithms shows higher performance than its topic-sensitive variant based on BLEU and ROUGE metrics.
Further investigation of TextRank and its topic-sensitive variant, we observed that topic-sensitive TextRank provides superior keywords when all query tokens occur in the document to be explained. When the document contains only one or a subset of query tokens, topic-sensitive PageRank leads to lower weights for document tokens related to non-present query tokens in the document compared to those by the original PageRank algorithm, and this hurts the quality of explanations for such cases by  topic-sensitive TextRank.

LIME and Sensitivity methods underperform the GenEx model where the performance difference is considerable and statistically significant. 
These models extract document tokens that highly impact the relevance probability of a document with respect to a query. Consequently, tokens that are exact match or semantically similar to query tokens, get the highest importance scores.
While this type of explanation show document relevance in terms of keyword matching, they mostly fail to provide words related to query aspects in their top keywords. 
These models thus provide different but complementary explanation than the GenEx model.
LIME explanation of  the TF.IDF ranker achieves better performance than that of  the BERT-based ranker over the Wiki and Clue-Res test sets. The TF.IDF and BERT-based rankers are used to get the relevance probability of perturbed documents. The lower performance of LIME for the BERT-based ranker could be related to the complex structure of BERT, where the widely used explanation based on linear approximation is not accurate enough to predicate its function. 

GenEx significantly outperforms KeyBERT that shows explanations of interest are beyond finding important noun phrases of a text segment.

Tables~\ref{table:perf-ROUGE} and~\ref{table:perf-BLEU} report the performance of the ablations of the GenEx model. 
As the results show, all model variants almost always outperform the original transformer model. The obtained improvements are also mostly statistically significant, with exceptional cases of  the Bleu-2/R-2 performance. This observation demonstrates the effectiveness of adding \emph{query-attention} layers and \emph{masked-query} decoding.

\texttt{Seg-q-toks} and \texttt{Sep-q-doc} models show similar performance on test sets, however our analysis of their generated explanations reveals that the two model perform well on almost disjoint sets of query-document pairs. 
Comparing the generated explanations by these two models, the main difference seems to arise from query ambiguity. 
The \texttt{Seg-q-toks} model generates better explanations for ambiguous queries than the \texttt{Sep-q-doc} model. 
One example of ambiguous queries in the Clue-Res dataset is query 73 of TREC Web track 2010, ``the sun'', with documents about  the star in solar system, the U.K. newspaper, and the Baltimore Sun newspaper.
In case of ambiguous queries, we believe that it would be helpful to encode queries using document tokens as their contexts. That is likely the main reason that the \texttt{Seg-q-toks} model outperforms the \texttt{Sep-q-doc} model for ambiguous queries.

\subsection{Semantic Evaluation of Explanations}
To semantically compare the generated explanations with respect to gold explanations, we use the BERTScore metric. This evaluation is necessary for abstractive text generation as BLEU and ROUGE metrics only consider the exact matching between generated and gold text segments, while a concept can be expressed in different ways.
The default setting of BERTScore, ``roberta-large\_L17\_no-idf\_version = 0.3.2 (hug\_trans = 2.8.0)-rescaled'', is used in our evaluations. 
The results are shown in Table~\ref{table:perf-Bertscore}. 
GenEx outperforms the original Transformer by 9.9\%, 10.7\%, and 23.2\% over \texttt{Wiki}, \texttt{Clue-Res}, and \texttt{Passex} datasets, respectively, and all improvements are statistically significant. These results demonstrate the higher quality of generated explanations by GenEx compared to those by the original Transformer. As both models are trained on the same data, these improvements also demonstrate the suitability of GenEx architecture for the task in hand.

The semantic evaluation results over \texttt{Clue-Res} and \texttt{Passex} datasets are almost consistent with those of Rouge and BLEU metrics; GenEx greatly outperforms the original transformer and its variants, and is not the best performing model variant on the \texttt{Wiki} test set. 
Although GenEx is not the best performing variant on the \texttt{Wiki} test set, it still outperforms the original Transformer and the improvements are statistically significant. 
The difference between model variants could be related to the special structure of Wikipedia articles that is not the case for all pages in the Web.  As the goal of our study is to explain a  relevant document  to a query, a model that is not dependent on the structure of document's content is more desirable. Thus, GenEx is designed and trained to perform well on general documents in \texttt{Clue-Res} and \texttt{Passex} without being trained on them. 

Explanations of keyword-based baselines are not evaluated using this metric as the BERT representations of input sequences in BERTScore are sensitive to input grammaticality and the order of words.

\begin{table*}
	\caption{Comparing GenEx, TextRank, and LIME-TF.IDF explanations based on human evaluation.}
	\label{table:annotation-genex-lime-tfidf}
	\centering 
	\begin{tabular}{ c c  c  c}
	\hline
    \multirow{2}{*}{$\mathcal{K}_f$} & \multicolumn{2}{c}{Majority prefer} & \multirow{2}{*}{$\frac{\mathrm{\# individual \; prefer\;(GenEx)}}{\mathrm{\# individual \; prefer\;(TextRank)\;}}$}  \\ \cline{2-3}
    & GenEx &  TextRank & \\ \hline
    0.50 & 73\% & 20\% &  3.3\\
\hline
	\end{tabular}
\end{table*}
\begin{table*}
		\begin{tabular}{ c c  c  c}
	\hline
    \multirow{2}{*}{$\mathcal{K}_f$} & \multicolumn{2}{c}{Majority prefer} & \multirow{2}{*}{$\frac{\mathrm{\# individual \; prefer\;(GenEx)}}{\mathrm{\# individual \; prefer\;(LIME)\;}}$}  \\ \cline{2-3}
    & $\mathrm{GenEx}$ &  $\mathrm{LIME}$ & \\ \hline
    0.37 & 50\% & 32\% &  2.3\\
\hline
	\end{tabular}
\end{table*}

\begin{table}
	\caption{Quality of GenEx explanations.	}
	\label{table:annotation-genex-quality}
	\centering 
	\begin{tabular}{ c c  c c c }
	\cline{2-4}
	& Grammaticality & Relevance to query & Relevanct to document  \\ \hline
	Avg. score & 4.06 & 3.23 & 3.07 \\ \hline
    $\mathcal{K}_f$ (binary) & 0.49 & 0.42 & 0.51 \\ 
\hline
	\end{tabular}
\end{table}

\subsection{Example Generated Explanations}
Table~\ref{table:example} shows two examples of explanations generated by the GenEx model. 
Document contents are shown in Table~\ref{table:doc_content_examples}. 
In addition to the generated explanations, the table shows the document titles and the top 3 keywords of documents obtained by the TextRank and LIME+TF.IDF models. Document titles show the general topic of the documents, which are not necessarily good explanations for queries that seek to find these documents. 
The GenEx model generates a reasonable explanation for the first example in the table. 
As this example shows query-focused explanation of the retrieved document is different than the document title which can be considered as  general explanation of the document. The generated explanation for example 1 is more useful than the document title to reveal the relevancy of the document to the query.
Note that document titles are not used as input of explanation generation models. Although it may be helpful to have document titles for contextual representation of documents, we do not want the model to rely on a feature that may not be available for some webpages and reduces the applicability of the model in practice.

Second example in Table~\ref{table:example} shows a failure of the GenEx model, where the generated explanation is part of the expansion of the given abbreviated query. A reasonable explanation for the second document with respect to the given query can be ``symptoms". 
Desired explanations to be generated by a model should not repeat query  terms as  long as  explanation readability is not sacrificed. 
This is the reason that masked query decoding is proposed in our GenEx model. 
However, 
expansion tokens of abbreviated queries in documents have high similarity values with query tokens, and they are not masked during the decoding since query masking is done based on exact matching of tokens. This input characteristic makes it highly probable that the decoder generates expansion tokens of abbreviations as explanations.
Abbreviated queries  thus constitute one category of failure cases of our GenEx model.

\section{Human Evaluation}
\label{sec:humancheck}

In Section~\ref{sec:usability} we found that people could use \textsc{GenEx}-style synthetic explanations to more accurately and more efficiently identify relevant documents. In Section~\ref{sec:exp} we evaluated the effectiveness of \textsc{GenEx} explanations using automatic evaluations. We now close the loop by exploring whether the automatically generated explanations serve to help people as the initial study suggested. The GenEx model as well as existing baseline models explain a single document with respect to a query, and thus the user studies in this section are designed according to the nature of models, and are different than the one conducted in Section~\ref{sec:usability}. We leave the explanation of the entire search results for future work.

We randomly selected 25 query-document pairs from each of the Clue-Res, Passex, and Wiki test sets (Section~\ref{sec:exp}).
For Clue-Res, we had the additional requirements that the documents must 
have a maximum length of 600 tokens (for ease of human review). 
In this study\footnote{Institutional Review Board number 1381.},
performed on Amazon Mechanical Turk,  we presented each worker with a query-document pair and asked questions addressing preference between explanations of different approaches, their linguistic quality, and their relevance to the query and document pair. The worker is shown two illustrative examples for orientation. Each query-document pair is evaluated by three different master workers located in the United States to reduce subjectivity.

\subsection{Comparison with Baseline Models}
For each query-document pair, workers are asked to answer whether: (1)~GenEx explanation is better, (2)~Explanation by a baseline (LIME-TF.IDF or TextRank) is better, (3)~both are equally good, or (4)~both are equally bad.

TextRank is mostly the best performing baseline in terms of BLEU and ROUGE metrics according to the results in Table~\ref{table:baseline-kw}.
LIME is a successful and  widely-used model for explanation of  black-box models, which  is also the state-of-the-art model in the explanation of black-box rankers~\cite{Singh:2019:EES:3289600.3290620}.
LIME has shown a better performance in explaining the TF.IDF ranker in our experiments with results reported in Table~\ref{table:baseline-kw}. LIME does not use any information about the structure of the TF.IDF ranker, it only uses its outputs for the perturbed instances of the document to be explained.
This baseline is chosen as GenEx explains why a document is relevant to a query without having any knowledge of the underlying ranker model.

Table~\ref{table:annotation-genex-lime-tfidf} summarizes the obtained results which demonstrate that the majority of workers preferred  explanations generated by GenEx over those by LIME-TF.IDF and TextRank in 18\%-53\% more samples, respectively. 
In addition, the ratios of individual workers who preferred GenEx explanations over LIME-TF.IDF and TextRank are 2.3 and 3.3, respectively. Results of human evaluation show the higher quality of explanations generated by GenEx compared to baseline models.

\subsection{Explanations Quality}
We also conducted a user study to capture human evaluation of quality of the GenEx explanations. Three different dimensions are considered for this evaluation: linguistic quality, relevance of explanation to the query topic, and relevance of explanation to the document content. The last two questions provide a proxy for the utility of generated explanations' ability to connect the query and document. 
We use the 5-point Likert Scale to evaluate the subjective quality of generated explanations, as Very poor, Poor, Acceptable, Good, and Very good, with assigned scores from 1 to 5, respectively. 
Because differences between the levels can be subtle, we calculated Fleiss' Kappa agreement by collapsing negative scores (1 and 2) to \emph{no} and the others to \emph{yes}. 
 Table~\ref{table:annotation-genex-quality} summarizes the obtained results.

 \textbf{Linguistic quality.} We  evaluated the grammaticality and coherence of the explanations generated by the GenEx model. 
As shown in Table~\ref{table:annotation-genex-quality}, we observed that the average of grammaticality scores is 4.06. 
Only 9\% of all answers to the grammaticality questions were Poor or Very poor. Part of this strong result can be due to the short length of explanations, as all generated explanations are terse (up to four terms). Yet as explanations are generated in an abstractive way, this result strongly indicates the potential of our GenEx model in generation relevance explanations.

\textbf{Relevance to query and document pairs.}
We asked workers if the generated explanations were relevant to the query topic and document content. 
Note that, following annotation guidelines,
explanations that do not provide additional information compared to the query, such as explanations that are a subsequence of queries, should be rated as very poor for both questions. 
An explanation that is relevant to the topic of the query without repeating the query, provides additional useful information with respect to the query that can help users in understanding the information space.  
When an explanation is relevant to the document, it means that the explanation is describing the content of the document, and the model is not generating a general or a high-frequent phrase in the training data, which is a common issue of text generation models.

The obtained results in Table~\ref{table:annotation-genex-quality} show that on average, workers rated GenEx explanations have acceptable degree of relevance to both query topic without repeating the query and document content. These results demonstrate that generated explanations can reasonably describe document content with respect to the query topic and provide more information than the given query.

We chose not to run this part of the study for LIME explanations, since by their nature, they are unlikely to be grammatical. On the other hand, 
extractive explanations by LIME are likely to be relevant to the document, and do not require the same evaluations as abstractive explanations by GenEx. Note that abstractive generation of texts at the level of noun-phrase has a lower risk of topic drift compared to abstractive snippet generation which has been motivated recently~\cite{Chen:2020}, and is confirmed by the results of human evaluation on relevance to document content.

\section{Conclusions and Future Work}

We studied how a retrieved document can be explained with respect to the given query. We proposed a Transformer-based  architecture with \emph{query attention} layers and \emph{masked-query} decoding, called \texttt{GenEx}. 
The proposed solution is not trained using any manually labeled training data. Comprehensive evaluation of GenEx demonstrated its superior performance.

We believe that this work opens up new directions towards explainable document retrieval. 
In the recent search scenarios with limited bandwidth interfaces, such as conversational search systems using speech-only or small-screen devices~\cite{Radlinski:2017}, presenting result lists with long snippets is not plausible, emphasizing the need for a new form of explanation conforming to their characteristics. 
We intend to incorporate  explanation into conversational search systems  with limited bandwidth interfaces.
Another interesting direction to pursue is to design an end-to-end explanation model that can handle documents of any length. 
Given the memory constraints of current hardware, GenEx works on query-biased extraction of long documents which may not be optimal. 
We also would like to extend GenEx to make it more robust with respect to diverse types of queries in Web search. 
Furthermore, the proposed solution generates an explanation for a query-document pair. Future work can explore document-level explanation based on the top results.
Incorporating the generated explanations on a web search interface in order to improve search experience for users is another interesting future direction.

\begin{acks}
This work was supported in part by the Center for Intelligent Information Retrieval, in part by Amazon.com, in part by NSF grant number 1813662 and in part by NSF grant \#IIS-2039449. Any opinions, findings and conclusions or recommendations expressed in this material are those of the authors and do not necessarily reflect those of the sponsor.
\end{acks}

\bibliographystyle{ACM-Reference-Format}
\bibliography{ref}


\end{document}